\documentclass{svproc}
\usepackage{url}
\usepackage{cite}
\usepackage{amsmath,amssymb,amsfonts}
\usepackage{algorithmic}
\usepackage{graphicx}
\usepackage{textcomp}
\usepackage{xcolor}
\usepackage{float}
\usepackage{multirow}

\begin{document}
\mainmatter              % start of a contribution
\title{Measuring Social Media Polarization Using Large Language Models and Heuristic Rules}
\titlerunning{Measuring Polarization with LLMs and Heuristics}  % abbreviated title (for running head)
%                                     also used for the TOC unless
%                                     \toctitle is used
%
\author{Jawad Chowdhury \and Rezaur Rashid \and Gabriel Terejanu}
\authorrunning{Chowdhury et al.} % abbreviated author list (for running head)
%
%%%% list of authors for the TOC (use if author list has to be modified)
\tocauthor{}
\institute{
Department of Computer Science, University of North Carolina at Charlotte, Charlotte, NC 28223, USA \\
\email{\{mchowdh5,\,mrashid1,\,gabriel.terejanu\}@charlotte.edu}
}

\maketitle              % typeset the title of the contribution

\begin{abstract}
Understanding affective polarization in online discourse is crucial for evaluating the societal impact of social media interactions. This study presents a novel framework that leverages large language models (LLMs) and domain-informed heuristics to systematically analyze and quantify affective polarization in discussions on divisive topics such as climate change and gun control. Unlike most prior approaches that relied on sentiment analysis or predefined classifiers, our method integrates LLMs to extract stance, affective tone, and agreement patterns from large-scale social media discussions. We then apply a rule-based scoring system capable of quantifying affective polarization even in small conversations consisting of single interactions, based on stance alignment, emotional content, and interaction dynamics. Our analysis reveals distinct polarization patterns that are event dependent: (i) anticipation-driven polarization, where extreme polarization escalates before well-publicized events, and (ii) reactive polarization, where intense affective polarization spikes immediately after sudden, high-impact events. By combining AI-driven content annotation with domain-informed scoring, our framework offers a scalable and interpretable approach to measuring affective polarization. The source code is publicly available at: \url{https://github.com/hasanjawad001/llm-social-media-polarization}
%%
% We would like to encourage you to list your keywords within
% the abstract section using the \keywords{...} command.
\keywords{Affective Polarization, Social Media Discourse, Large Language Models, Stance Detection, AI for Social Impact}
\end{abstract}

%%%%%%%%%%%%%%%%%%%%%%%%%%%%%%%%%%%%%%%%%%%%%%%%%%%%%%%%%%%%%%%%%%%
\section{Introduction}
\label{sec:int}
The rise of social media platforms in recent years has transformed political discourse by enabling real-time information exchange and broader audience engagement~\cite{garimella2018political, saaida2023role}. This transformation, driven by the evolving media landscape, continues to shape how information is produced, distributed, and consumed while simultaneously redefining how individuals interact and maintain connections in digital spaces~\cite{marlowe2017digital, dahlgren2009media, chambers2013social, cinelli2021echo}. While these platforms facilitate engagement, they have also intensified ideological divisions, as algorithmic content curation reinforces preexisting beliefs by prioritizing content aligned with users’ prior views, limiting exposure to diverse perspectives. This selective exposure contributes to \textit{affective polarization}, where individuals develop strong positive emotions toward their in-group members while exhibiting hostility toward those from opposing groups or with opposing views~\cite{feldman2023affective, serrano2021digital, overgaard2024perceiving}. Studies suggest that such polarization is not only shaped by political ideology but also by the emotional tone, discourse structure, and interaction patterns within online discussions. Affective polarization has been linked to increased political radicalization, reduced bipartisan cooperation, and the spread of misinformation~\cite{yair2020note, yu2024}. Understanding its dynamics is crucial for evaluating the broader societal implications of online discourse.

Social media platforms, such as Twitter (now X), can further amplify these divisions through algorithmic content curation, which prioritizes engagement-driven interactions and often promotes sensationalized, polarizing content~\cite{cinelli2021echo, rodilosso2024filter, piccardi2024social}. Research suggests that online echo chambers reinforce polarization by predominantly exposing users to like-minded perspectives while restricting interaction with opposing viewpoints~\cite{gillani2018me, iyengar2019origins}. However, while exposure to counter-ideological content has the potential to correct misperceptions and reduce polarization in some cases~\cite{druckman2022mis}, it can also provoke defensive responses, particularly in contentious social movements, where ideological conflict is often accompanied by toxic interactions and digital aggression~\cite{suarez2022toxic}. These dynamics highlight the complex role of social media in shaping ideological divides and underscore the need for robust methodologies to quantify and analyze affective polarization in online discourse.

%%
% Existing approaches to measuring affective polarization in social media largely rely on sentiment analysis \cite{tyagi2020affective}, stance detection, and network-based polarization indices \cite{guerra2013measure}. While these methods provide valuable insights, they often struggle with linguistic nuances, context-dependent stance shifts, and the dynamic nature of online discourse. Recently, large language models (LLMs) have emerged as powerful tools for text classification and content analysis, offering more sophisticated natural language understanding than traditional sentiment classifiers \cite{brown2020language}. Studies have leveraged LLMs for detecting political bias, misinformation \cite{pendyala2024explaining}, and ideological framing, but their application in systematically quantifying affective polarization remains underexplored.
%%
Existing approaches to measuring affective polarization in social media largely rely on sentiment analysis, stance detection, and/or network-based polarization indices~\cite{tyagi2020affective, guerra2013measure, rashid2024quantifying, martinez2024methodology}. Sentiment analysis techniques classify text as positive, negative, or neutral, providing a general sense of emotional tone but often failing to capture the complexity of political discourse, such as sarcasm or implicit bias~\cite{rani2024comprehensive}. Stance detection methods aim to determine whether a user supports, opposes, or remains neutral on an issue, yet they frequently struggle with linguistic nuances, especially in highly polarized debates where positions are subtly framed~\cite{aldayel2021stance, mets2024automated}.  Recent studies have combined multimodal signals~\cite{thareja2024multimodal} or social network structures~\cite{yarchi2024political, lerman2024affective} to improve measurement, but these still face limitations in capturing the nuanced emotional dimensions driving polarization.
% Thareja~\cite{thareja2024multimodal} developed a multimodal framework combining text and image embeddings for sentiment analysis on social media, improving detection of extreme sentiments, but it does not address polarization dynamics. Yarchi et al.~\cite{yarchi2024political} analyzed affective polarization across platforms, capturing interactional patterns, yet relied on traditional sentiment methods limited in handling nuanced emotional contexts. Similarly, network-based polarization indices analyze user interactions and retweet behaviors to map ideological divides. Feldman et al.~\cite{lerman2024affective} leveraged social network data and affective signals to map polarization dynamics, offering insights into interaction patterns, but their approach lacks integration of contextual text analysis for nuanced emotional scoring, highlighting that these methods often overlook the affective dimension, namely the role of emotional intensity in shaping polarization.

Recently, large language models (LLMs) have emerged as powerful tools for text classification and content analysis, offering deeper contextual understanding than traditional sentiment classifiers~\cite{brown2020language, chae2023large, zhang2023sentiment}. LLMs leverage vast amounts of data to better capture subtle variations in ideological framing, rhetorical strategies, and the emotional undercurrents of online discourse~\cite{wang2023emotional, khan2025leveraging}. Their ability to process language in a more context-aware manner makes them particularly well-suited for analyzing sentiment shifts, emotional intensity, and therefore affective polarization. Studies have successfully applied LLMs for detecting political bias, misinformation, and ideological framing, demonstrating their potential in large-scale social media analysis~\cite{pendyala2024explaining, linegar2023large, zhang2024toward}. However, despite these advancements, the application of LLMs in systematically quantifying affective polarization, which involves measuring both stance alignment and emotional intensity, remains underexplored. Addressing this gap requires a synergetic approach that can combine LLM-driven annotation with domain expertise to ensure a more interpretable and scalable approach to measuring affective polarization across social media platforms.

Building on this line of work, our study introduces a significant departure from existing methodologies. Rashid et al. \cite{rashid2024quantifying} developed a counterfactual framework that leverages the network-based polarization index~\cite{tyagi2020affective} to examine the role of influential users in shaping polarization on Twitter (now X). Their study assessed how removing influential conversations altered polarization scores but relied on pre-defined sentiment-based classifiers. In contrast, our work employs LLM-based annotation to extract stance, affect, and agreement patterns from large-scale discussions, enabling a more nuanced understanding of affective polarization. 

Furthermore, we integrate heuristic rules to score polarization, addressing limitations in existing sentiment-based approaches and ensuring interpretability in measuring discourse intensity. Specifically, our LLM-driven framework classifies tweets based on their stance (support, opposition, or neutrality on an issue), affective content (presence of emotionally charged language indicative of polarization), and agreement patterns (the extent to which replies align or conflict with the stance of the original post). These extracted attributes are then used within a structured scoring system that quantifies polarization by evaluating stance alignment, emotional intensity, and disagreement dynamics. 

\begin{figure}[H]
    \centerline{\includegraphics[width=0.99\linewidth]{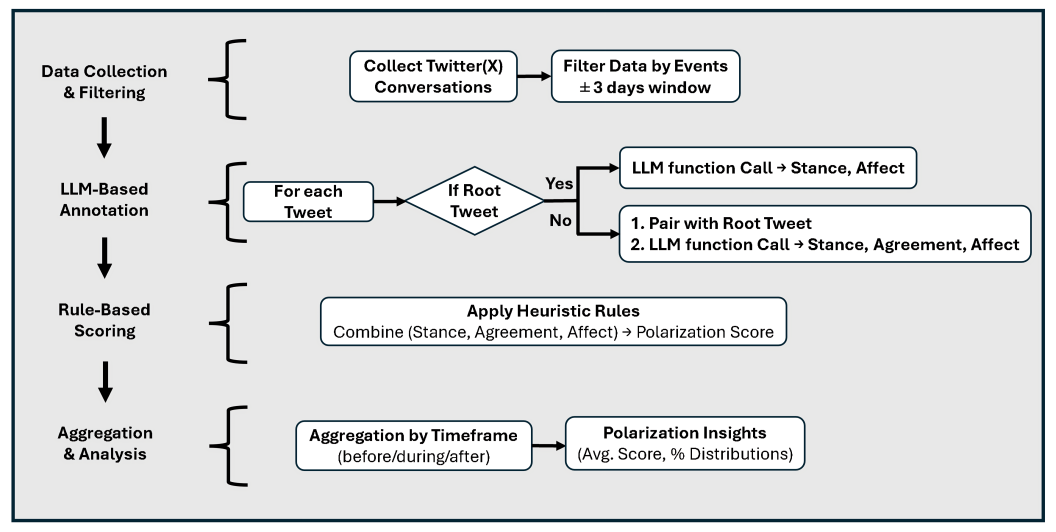}}
    \caption{Detailed workflow pipeline of our methodology, illustrating data collection and filtering, LLM-based annotation, application of heuristic rules, and aggregation for polarization insight generation.}
    \label{fig:met_pipeline}
    % \vspace{-0.4in}
\end{figure}

By leveraging LLMs and domain-informed heuristics, our method provides a scalable, interpretable, and more context-aware approach to polarization measurement than traditional sentiment-based techniques. The major contributions of our study are stated as follows:
\begin{itemize}
\item We propose a novel LLM-based framework for measuring affective polarization, enhancing traditional sentiment analysis-based methods by incorporating large-scale language understanding.
\item We introduce a scoring system that systematically quantifies affective polarization, capturing nuanced discourse dynamics such as stance shifts, emotional intensity, and disagreement patterns. Unlike network-based polarization measures, our scoring system is explainable and can effectively quantify polarization even in very small conversations involving only a single interaction.
\item We conduct a large-scale empirical analysis of affective polarization in highly contentious discussions on climate change and gun control, uncovering key event-driven polarization trends and distinguishing anticipatory vs. reactive polarization.
\end{itemize}
% \begin{itemize}
%     \item We propose a novel LLM-based framework for measuring affective polarization, enhancing traditional sentiment analysis-based methods by incorporating large-scale language understanding.
%     \item We introduce an expert-defined heuristic scoring system that systematically quantifies affective polarization, capturing nuanced discourse dynamics such as stance shifts, emotional intensity, and disagreement patterns.
%     \item We conduct a large-scale empirical analysis of affective polarization in highly contentious discussions on climate change and gun control, uncovering key event-driven polarization trends and distinguishing anticipatory vs. reactive polarization.
% \end{itemize}

The remainder of this paper is structured as follows: Section~\ref{sec:met} details our proposed framework, covering data collection, LLM-based annotation, and affective polarization scoring using heuristic rules. Section~\ref{sec:res} presents our findings, analyzing how affective polarization evolves over time in response to major events. Finally, Section~\ref{sec:con} offers concluding remarks and outlines directions for future research.
\section{Proposed Approach \& Implementation}
\label{sec:met}
This study employs large language models (LLMs) to analyze affective polarization in online discussions surrounding contentious issues. Unlike previous studies that relied on sentiment classifiers or rule-based stance detection, we introduce a hybrid approach that integrates opensourced pretrained large language model LLaMA 3.1 70B for automatic text analysis with predefined scoring system to systematically quantify affective polarization. Our approach leverages LLMs to extract critical attributes such as stance, affective content, and agreement levels between users and their posts, while predefined rules assign affective polarization scores based on interaction patterns.

The overall workflow pipeline followed in our proposed work is illustrated in Figure~\ref{fig:met_pipeline}, and consists of four main stages. \textbf{(1) Data collection:} retrieving Twitter (X) conversation threads related to highly debated sociopolitical topics e.g. climate change and gun control. \textbf{(2) LLM-based annotation:} extracting stance, affect, and agreement between tweets using LLaMA 3.1 70B. \textbf{(2) Predefined rule application:} assigning affective polarization scores using domain heuristics based on stance alignment, affect, and agreement information extracted by LLM in the previous step. \textbf{(4) Aggregation \& analysis:} computing polarization score at the conversation level and evaluating polarization trends over time.

%%%%%%%%%%%%%%%%%%%%%%%%%%%%%%%%%%%%%%%%%%%%%%%%%%%%%%%%%%%%%%%%%%%
\subsection{Data Collection and Structure}
Our study employs a comprehensive dataset from Twitter (now X), focusing on two contentious political issues - climate change and gun control using a keyword-based approach. An initial set of tweets were retrieved via the Twitter API prior to its 2023 restrictions; based on curated keywords and hashtags relevant to each topic, including terms like “\#ClimateCrisis,” “\#GunReformNow,” and event-specific phrases tied to major incidents (e.g., IPCC reports, mass shootings). To capture full user interactions, conversation cascades were expanded recursively to include all referenced tweets (replies, quotes, parents), ensuring inclusion of relevant discussions even without explicit keyword matches.

The climate change dataset spans June 1, 2021, to May 31, 2022, comprising 46M tweets from 4.8M unique users across 726,378 conversation threads of at least three tweets. The gun control dataset covers January 1, 2022, to December 31, 2022, with 14.4M tweets from 2.66M unique users across 335,000 conversation threads. To focus on threads with substantial engagement and to mitigate the influence of trivial or low-activity conversations, threads were restricted to those with $\ge$ 20 tweets and $\ge$ 10 unique users. Finally, each dataset consists of structured conversation threads, which we define as:
\begin{itemize}
    \item \textbf{Parent Tweet:} The original post initiating the discussion.
    \item \textbf{Child Tweet:} Replies engaging with the parent tweet.
\end{itemize}

Each tweet-reply pair was analyzed independently to extract stance, affect, and agreement, forming the basis for our affective polarization scoring. While the dataset contains a broad collection of conversations, our analysis specifically focuses on eight key events: four related to climate change and four to gun control. For each event, conversations were segmented into three distinct timeframes:
\begin{itemize}
    \item \textbf{Before:} Conversations occurring from 3 days before the event up to the day before the event starts.
    \item \textbf{During:} Conversations occurring between the official start and end dates of the event.
    \item \textbf{After:} Conversations occurring from the day after the event ends to 3 days after.
\end{itemize}

This segmentation allows us to analyze how affective polarization evolves over time, particularly whether polarization intensifies in anticipation of an event (before), peaks during the event (during), or escalates in response to its aftermath (after). The selection of these events is detailed in Section \ref{sec:res}. By structuring our dataset with these temporal segments, we provide a granular view of online discourse dynamics, enabling comparisons across different event types and their corresponding shifts in polarization levels.

%%%%%%%%%%%%%%%%%%%%%%%%%%%%%%%%%%%%%%%%%%%%%%%%%%%%%%%%%%%%%%%%%%%
\subsection{LLM-Based Classification}
We employed LLaMA 3.1 70B, an open-source pretrained model specialized in understanding and generating human-like text with structured prompt engineering to classify:
\begin{itemize}
    \item \textbf{Tweet Stance:} \textit{Belief, Disbelief, Do Not Know} (for climate change); \textit{Pro, Anti, Do Not Know} (for gun control).
    \item \textbf{Tweet Affect:} Whether the tweet contains emotionally charged language indicative of affective polarization.
    \item \textbf{Agreement Level:} Whether the child tweet agrees or disagrees with the parent tweet.
\end{itemize}
To ensure deterministic outputs, we configured the model to minimize randomness in classification by setting the temperature of the LLM to zero.

%%%%%%%%%%%%%%%%%%%%%%%%%%%%%%%%%%%%%%%%%%%%%%%%%%%%%%%%%%%%%%%%%%%
\subsection{Affective Polarization Scoring with Heuristic Rules}
Building upon the LLM-extracted attributes from the previous section, we define the affective polarization score using heuristic rules. These rules account for critical aspects of interaction dynamics between the parent tweet (original post) and the child tweet (reply), considering three key factors: stance alignment, which determines whether the reply tweet aligns or opposes the stance of the parent tweet; affective expression, assessing whether any of the tweets exhibit strong emotional language indicative of polarization; and agreement level, evaluating whether the reply tweet explicitly agrees or disagrees with the parent tweet. By combining LLM-based classification with structured heuristic rules, we enable a systematic, scalable, and interpretable quantification of affective polarization across social media discourse.

In our scoring system in Table~\ref{tab:met_heu}, low scores (0 to 4) reflect constructive interactions and respectful disagreements. Specifically, a score of 0 represents the ideal, highlighting civil exchanges that actively seek mutual understanding and collaboration despite differing views, fostering openness and healthy dialogue. Conversely, high scores (8 and 10) represent interactions dominated by affective polarization, including heated conflict, incivility, and emotional hostility. A score of 10 is especially concerning, indicating interactions entirely within ideological echo chambers that reinforce negative emotions without exposure to opposing perspectives, thus intensifying polarization and potentially driving further hostility. High polarization scores, particularly in the upper half of the spectrum in Table~\ref{tab:met_heu}, represent a shift away from productive dialogue toward emotional reactivity and out-group animosity, undermining effective communication and reinforcing intolerance. The stark contrast between scores 0 and 10 emphasizes the importance of fostering respectful, diverse discourse.

Importantly, our proposed heuristic scoring approach provides a significant advantage over traditional statistical methods by effectively quantifying polarization even in small conversations consisting of a single interaction. The overall polarization score for longer conversations is then computed by averaging the polarization scores across all individual interactions, ensuring scalability and adaptability to varied conversation lengths and structures.

\begin{table*}[htpb]
\renewcommand{\arraystretch}{1.5}
\caption{Heuristic Rules for Polarization Scoring. The table defines interaction categories based on stance similarity, affect presence, and agreement between tweets.}
\begin{center}
\resizebox{\textwidth}{!}{
\begin{tabular}{|p{2.7cm}|p{2.5cm}|p{1.9cm}|c|p{7.2cm}|}
\hline
\textbf{Reply vs. Parent Stance} & \textbf{Affect in Reply or Parent} & \textbf{Agreement} & \textbf{Score} & \textbf{Discourse Quality Category and Description} \\
\hline
Opposite Stance & No (in both) & Yes & 0 & \textbf{Constructive Dialogue} - Constructive discussion with no negative affect and mutual agreement.\\
\hline
Same Stance & No (in both) & No & 2 & \textbf{Cordial Disagreement} - Healthy debate within the same stance group.\\
\hline
Opposite Stance & No (in both) & No & 4 & \textbf{Respectful Disagreement} - Polite disagreement across opposing stance groups without negative emotions.\\
\hline
Same Stance & No (in both) & Yes & 6 & \textbf{Echoic Agreement} - Echo chamber effect but without emotional hostility.\\
\hline
Opposite Stance & Yes (in either) & Yes & 6 & \textbf{Hard-Fought Agreement} - Agreement across opposing stances but with negative emotions.\\
\hline
Opposite Stance & Yes (in either) & No & 8 & \textbf{Heated Conflict} - Hostile disagreement between opposing stance groups.\\
\hline
Same Stance & Yes (in either) & No & 8 & \textbf{Discordant Allies} - Intra-group disagreement but with strong negative emotions.\\
\hline
Same Stance & Yes (in either) & Yes & 10 & \textbf{Polarizing Echo Chamber} - Echo chamber with emotional reinforcement, no interaction with opposing stance.\\
\hline
\end{tabular}
}
\label{tab:met_heu}
\end{center}
\end{table*}
\begin{table*}[htpb]
\renewcommand{\arraystretch}{1.5}
\caption{Illustrative example of LLM-based annotation. This table presents simulated tweets resembling real discussions to ensure anonymity, along with LLM-extracted attributes such as stance, affect, agreement information, and their explanations by the language model employed.}
\begin{center}
\resizebox{\textwidth}{!}{
\begin{tabular}{|p{3cm}|p{7cm}|p{7.5cm}|}
\hline
% \textbf{Attribute} & \textbf{Parent Tweet} & \textbf{Reply Tweet} \\
\multirow{2}{*}{\textbf{Attribute}} & \multirow{2}{*}{\textbf{Parent Tweet}} & \multirow{2}{*}{\textbf{Reply Tweet}} \\
& & \\ % Empty row for spacing
\hline
\textbf{Tweet Text} & 
`The science is clear. Climate change is real, and urgent action is needed before it's too late!' & 
`Climate change is a hoax pushed by the corrupt elites to brainwash the masses. Wake up, fool!'\\
\hline
\textbf{Stance} & \textbf{Belief} & \textbf{Disbelief} \\
% \hline
\textbf{Explanation} & 
The tweet clearly states that climate change is real, indicating a belief in the topic. & 
This tweet expresses disbelief in climate change, calling it a hoax and implying that those who believe in it are brainwashed. The tone is confrontational and dismissive. \\
\hline
\textbf{Affect} & \textbf{No} & \textbf{Yes} \\
% \hline
\textbf{Explanation} & 
Although the tone of the tweet is urgent and emphasizes the need for action, it does not contain deeply negative emotions or attitudes toward opposing views. & 
The tweet contains deeply negative emotions towards people who hold opposing views on climate change, calling them `fool' and implying they are brainwashed. This language is emotionally charged and divisive. \\
\hline
\textbf{Agreement} & \textbf{-} & \textbf{No} \\
% \hline
\textbf{Explanation} & - & 
The parent tweet and the reply have opposing views on climate change. The reply denies its existence, while the parent tweet affirms it and calls for urgent action. They clearly disagree on the topic. \\
\hline
\end{tabular}
}
\label{tab:met_exa}
\end{center}
\end{table*}

%%%%%%%%%%%%%%%%%%%%%%%%%%%%%%%%%%%%%%%%%%%%%%%%%%%%%%%%%%%%%%%%%%%
\subsection{An Illustrative Example: Prompting, Annotation, and Scoring}
To demonstrate the combined effect of LLM-based annotation and predefined scoring, we present an illustrative example using the LLaMA 3.1 70B model. 
Table~\ref{tab:met_exa} showcases the extracted stance, affect, and agreement attributes for a simulated conversation thread with just two tweets, while Table~\ref{tab:met_heu} illustrates the heuristic rules. The annotations in Table~\ref{tab:met_exa} were generated using a prompt template where the model was provided with paired tweets in the following format:
\begin{quote}\small\ttfamily
Tweet1: [Reply Tweet] \\
Tweet2: [Parent Tweet]
\end{quote}
In cases where Tweet1 was the parent tweet (i.e., the first tweet in a thread), Tweet2 was provided as an empty string (\texttt{""}), and agreement-related fields were not applicable. Along with this input, the model received the following function-call instructions:

\begin{quote}\small\ttfamily
Analyze the content of the provided tweets to assess their stance and emotional tone with respect to the topic: climate change. \\
Provide detailed classifications for stance, agreement between tweets, and affective polarization (emotional negativity towards opposing views).
\end{quote}

The function call requested the model to generate six fields, each with specific instructions:
\begin{itemize}
    \item \texttt{tweet1\_stance\_explanation}: 
    % \begin{quote}\small\ttfamily
    Provide a brief explanation of tweet1's stance on the topic climate change. If tweet1 expresses belief that climate change is real or expresses disbelief, explain the reasoning. If the stance is unclear, label it as don't know.
    % \end{quote}

    \item \texttt{tweet1\_stance}: 
    % \begin{quote}\small\ttfamily
    Classify tweet1's stance on the topic: climate change. Possible values: belief, disbelief, don't know. This classification should be based on the explanation provided in 'tweet1\_stance\_explanation'.
    % \end{quote}

    \item \texttt{tweets\_agreement\_explanation}: 
    % \begin{quote}\small\ttfamily
    Provide an explanation of whether tweet1 and tweet2 agree or disagree on the topic: climate change. Agreement indicates similar views; disagreement means opposing views. If tweet2 is not available, state `not applicable'. If the agreement is unclear, provide reasoning.
    % \end{quote}

    \item \texttt{tweets\_agreement}: 
    % \begin{quote}\small\ttfamily
    Classify the agreement between tweet1 and tweet2 with respect to the topic: climate change. Possible values: yes (agreement), no (disagreement), don't know (unclear).
    % \end{quote}

    \item \texttt{tweet1\_affect\_explanation}: 
    % \begin{quote}\small\ttfamily
    Explain whether tweet1 contains deeply negative emotions or attitudes specifically towards people who hold opposing views on the topic: climate change. The focus is on emotional negativity beyond the stance itself.
    % \end{quote}

    \item \texttt{tweet1\_affect}: 
    % \begin{quote}\small\ttfamily
    Classify tweet1's affective polarization, i.e., emotional negativity specifically towards opposing views on the topic: climate change. Possible values: yes (contains affective polarization), no (doesn't contain), don't know (uncertain).
    % \end{quote}
\end{itemize}

This prompt structure was applied consistently to generate the annotations presented in Table~\ref{tab:met_exa} and throughout our experiments. Based on the extracted attributes by the LLM-based annotations and applying the heuristic rules to the extracted features, the reply tweet is classified as Heated Conflict (Score = 8), as it: (1) holds the opposite stance, (2) contains emotionally charged language, and (3) expresses disagreement, indicating hostility between opposite stance groups. This example highlights a simple illustration of our approach and the ability of the framework to systematically evaluate and categorize affective polarization.

%%%%%%%%%%%%%%%%%%%%%%%%%%%%%%%%%%%%%%%%%%%%%%%%%%%%%%%%%%%%%%%%%%%
\subsection{Implementation Details}  
The classification pipeline was implemented using LLaMA 3.1 70B via Ollama for stance and affect classification, combined with Langchain for structured prompt engineering and deterministic response generation. Our initial dataset encompasses a total of $2,551$ conversations across the climate change-related events and $3,201$ conversations across the gun control-related events, considering different timeframes for each event. This hybrid approach provides an efficient and interpretable method for large-scale affective polarization quantification, bridging computational advancements in LLMs with human-guided analysis.

%%%%%%%%%%%%%%%%%%%%%%%%%%%%%%%%%%%%%%%%%%%%%%%%%%%%%%%%%%%%%%%%%%%
\section{Case Studies \& Results}
\label{sec:res}
To systematically analyze affective polarization on social media, we examined its evolution across two case studies as previously mentioned: climate change and gun control. Our framework quantifies polarization before, during, and after key events, allowing us to observe distinct temporal patterns in online discourse. The following subsections present findings for each case study, highlighting how different types of events influence the dynamics of affective polarization.

%%%%%%%%%%%%%%%%%%%%%%%%%%%%%%%%%%%%%%%%%%%%%%%%%%%%%%%%%%%%%%%%%%%
\subsection{Affective Polarization in Climate Change Discourse}
\begin{figure}[H]
    \centerline{\includegraphics[width=0.87\linewidth]{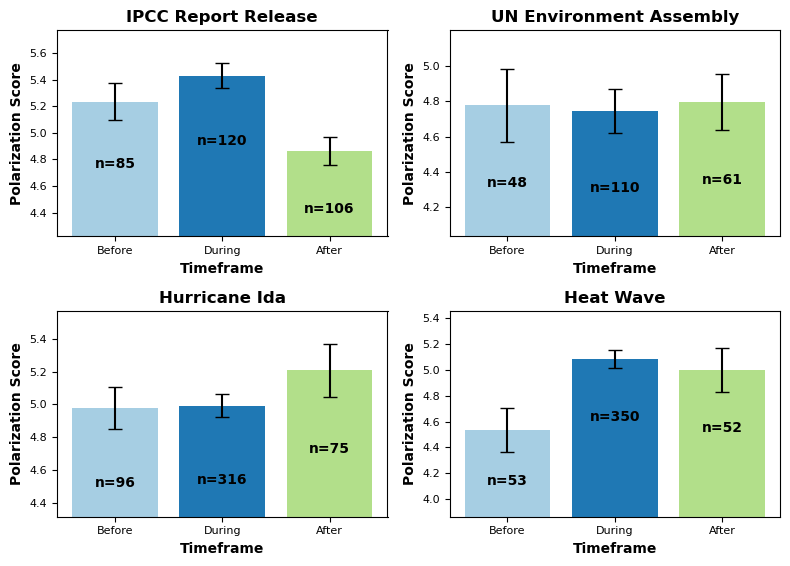}}
    \caption{Affective polarization scores across different timeframes (before, during, and after) for four climate change-related events. The bar plots illustrate the average polarization score, with error bars representing the standard error of the mean. The variable $n$ in each bar denotes the number of conversations analyzed during the corresponding timeframe for each event.}
    \label{fig:res_cc_pscore}
\end{figure}

In this section, we analyzed discussions surrounding major climate change-related events as one of the case studies for the current research. Specifically, we examined how affective polarization levels fluctuate before, during, and after four key climate-related events: the Intergovernmental Panel on Climate Change (IPCC) assessment report release, the United Nations (UN) Environment Assembly, Hurricane Ida and a major heatwave. To provide a structured overview, Table~\ref{tab:res_cc} presents the starting and ending timeframes of the considered events. For each event, we computed the mean affective polarization scores and the standard errors across all conversations before, during, and after the event. The results are illustrated in Figure~\ref{fig:res_cc_pscore}. As shown in the figure, affective polarization scores increase noticeably during and after these events compared to the preceding period. This aligns with expectations that major climate-related developments amplify emotional intensity in public discourse. Additionally, the volume of conversations (denoted by `\textit{n}’ in the figure) also significantly increases during and after these events, reflecting heightened social media interactions.

%As we can see from the figure, affective polarization scores exhibit a noticeable increase during and after these events compared to the preceding period. This trend aligns with expectations that major climate-related developments amplify engagement and emotional intensity in public discourse. Additionally, the volume of conversations (denoted by `\textit{n}' in the figure) also increases significantly during and after these events, further reflecting heightened social media interactions.

\begin{figure}[H]
    \centerline{\includegraphics[width=0.87\linewidth]{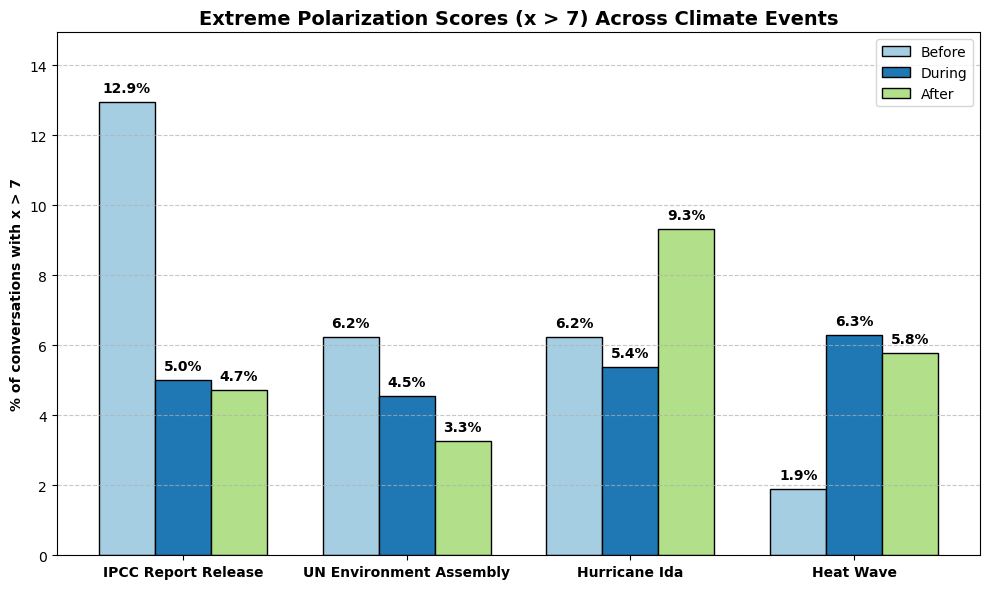}}
    \caption{Percentage of conversations exhibiting extreme polarization scores ($x > 7$) across different timeframes (before, during, and after) for climate change-related events.}
    % \caption{Distribution of polarization intensity in conversations across different timeframes (before, during, and after) for four climate change-related events. Conversations are categorized into three polarization levels: $x < 2.5$ (\textit{mild}), $2.5 \leq x \leq 7.5$ (\textit{moderate}), and $x > 7.5$ (\textit{extreme}). The bars represent the proportion of conversations within each category for each event.}
    \label{fig:res_cc_percentage}
\end{figure}

To better understand the intensity of polarization, we focused exclusively on conversations exhibiting extreme polarization (scores greater than 7). Figure~\ref{fig:res_cc_percentage} shows the percentage of highly polarized conversations during different timeframes (before, during, and after) for each climate-related event. The analysis reveals distinct patterns: well-publicized events such as the IPCC assessment report release and the UN Environment Assembly displayed heightened extreme polarization primarily before these events, reflecting anticipation and ideological engagement. Conversely, unpredictable climate events like Hurricane Ida and the heatwave experienced spikes in extreme polarization predominantly during and after the events, highlighting the role of spontaneous emotional responses and reactive discourse in shaping polarization patterns.
% To better understand the intensity distribution of polarization, we categorized all the conversations into three different levels based on the extremity of the affective polarization: \textit{low} ($x < 2.5$), \textit{moderate} ($2.5 \leq x \leq 7.5$), and \textit{high} ($x > 7.5$). Figure~\ref{fig:res_cc_percentage} illustrates the percentage of conversations in each category during different timeframes for each event. This illustration reveals that while most conversations exhibit moderate polarization levels, the prevalence of extreme polarization ($x > 7.5$) varies across different events. Well-publicized events such as the IPCC assessment report release and the UN environment assembly tend to show increased extreme polarization levels in conversations before and during the event, suggesting pre-event anticipation and heightened ideological engagement. In contrast, for more unpredictable climate disasters such as Hurricane Ida and the heatwave, extreme polarization primarily spikes during and after the event, indicating that spontaneous emotional reactions and reactive discourse shape polarization patterns.

\begin{table}[htbp]
\caption{Climate change-related events analyzed in this study, including their start and end dates.}
\begin{center}
\begin{tabular}{|c|c|c|}
\hline
\textbf{Event} & \textbf{Start Date} & \textbf{End Date} \\
\hline
IPCC Assessment Report Release & 2021-08-09 & 2021-08-09 \\
Hurricane Ida & 2021-08-26 & 2021-09-04 \\
Major Heatwave & 2021-06-25 & 2021-07-07 \\
UN Environment Assembly & 2022-02-28 & 2022-03-02 \\
\hline
\end{tabular}
\label{tab:res_cc}
\end{center}
\end{table}
% \vspace{-0.5in}

These findings suggest that the temporal characteristics of affective polarization in climate-related discussions are closely linked to the nature of the triggering event. Anticipatory polarization is more prominent for scheduled, policy-driven events, whereas reactive polarization is dominant for sudden climate-related disasters.

%%%%%%%%%%%%%%%%%%%%%%%%%%%%%%%%%%%%%%%%%%%%%%%%%%%%%%%%%%%%%%%%%%%
\subsection{Affective Polarization in Gun Control Discourse}
\begin{figure}[H]
    \centerline{\includegraphics[width=0.87\linewidth]{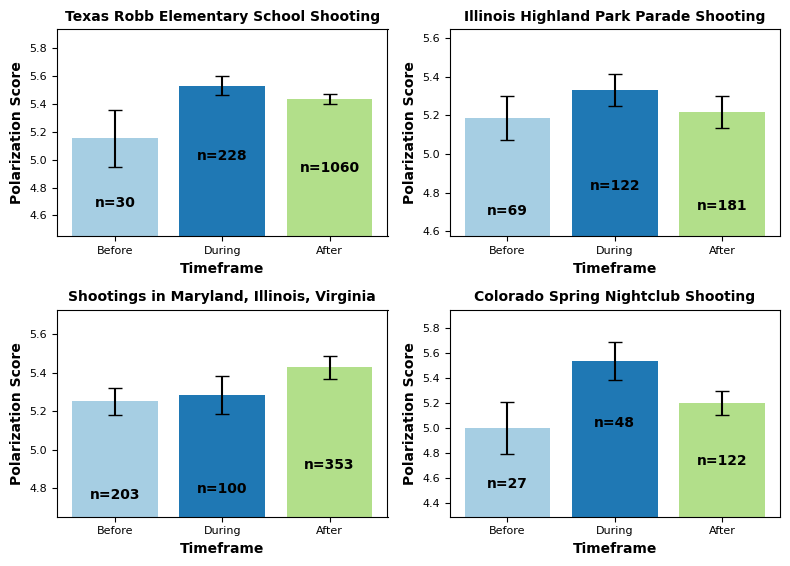}}
    \caption{Affective polarization scores across different timeframes (before, during, and after) for four gun control-related events. The bar plots display the average polarization score, with error bars indicating the standard error of the mean. The variable $n$ in each bar represents the number of conversations analyzed during the corresponding timeframe for each event.}
    \label{fig:res_gc_pscore}
\end{figure}

This section presents an analysis of discussions surrounding major gun control-related events to examine affective polarization trends in the aftermath of mass shootings. Specifically, similar to the previous case study, we analyzed how affective polarization fluctuated before, during, and after four major gun-related incidents: the Texas Robb Elementary School Shooting, the Illinois Highland Park Parade Shooting, Multiple Shootings in Maryland, Illinois, and Virginia, and the Colorado Spring Nightclub Shooting. Table~\ref{tab:res_gc} presents the starting and ending timeframes for the considered events. For each event, we computed the mean affective polarization scores and the standard errors across all social media conversations before, during, and after the event. As seen in Figure~\ref{fig:res_gc_pscore}, affective polarization scores increase significantly during and after these events compared to the period before. This pattern suggests that mass shootings amplify ideological divides in gun control discussions, with emotional intensity peaking n response to the event. The number of conversations (denoted by `\textit{n}’ in the figure) also substantially increases during and after these events, reflecting heightened online engagement. 

%This pattern suggests that mass shooting incidents strongly amplify ideological divides in gun control discussions, with emotional intensity peaking in response to the event. Additionally, in this case study, we also see the number of conversation (denoted by `\textit{n}' in the figure) increases substantially during and after these events, reflecting the heightened engagement in online discourse.

\begin{figure}[H]
    \centerline{\includegraphics[width=0.87\linewidth]{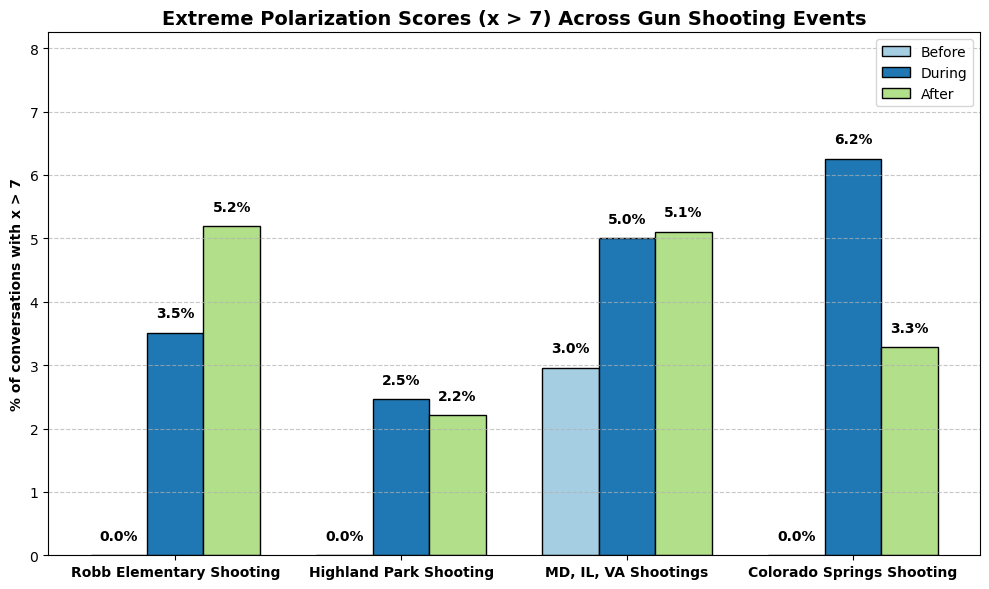}}
    \caption{Percentage of conversations exhibiting extreme polarization scores ($x > 7$) across different timeframes (before, during, and after) for gun control-related events.}
    % \caption{Distribution of polarization intensity in conversations across different timeframes (before, during, and after) for four gun control-related events. Conversations are categorized into three polarization levels: $x < 2.5$ (\textit{mild}), $2.5 \leq x \leq 7.5$ (\textit{moderate}), and $x > 7.5$ (\textit{extreme}). The bars indicate the proportion of conversations within each category for each event.}
    \label{fig:res_gc_percentage}
\end{figure}

\begin{table}[H]
\caption{Gun control-related events analyzed in this study, including the starting and ending dates.}
\begin{center}
\begin{tabular}{|c|c|c|}
\hline
\textbf{Event} & \textbf{Start Date} & \textbf{End Date} \\
\hline
Texas Robb Elementary School Shooting & 2022-05-24 & 2022-05-24 \\
Illinois Highland Park Parade Shooting & 2022-07-04 & 2022-07-04 \\
Multiple Shooting in Maryland, Illinois, Virginia & 2022-06-07 & 2022-06-07 \\
Colorado Spring Nightclub Shooting & 2022-11-19 & 2022-11-20 \\
\hline
\end{tabular}
\label{tab:res_gc}
\end{center}
\end{table}
% \vspace{-0.3in}

Similar to our previous analysis, we exclusively examined conversations with extreme polarization scores (greater than 7). Figure~\ref{fig:res_gc_percentage} illustrates the proportion of highly polarized conversations occurring before, during, and after four gun-related events. The analysis clearly indicates that extreme polarization is notably elevated during and after mass shooting incidents. Unlike climate change events, which often saw spikes in extreme polarization before anticipated events, gun control-related discussions primarily demonstrate reactive polarization, spiking directly in response to shootings. 
%This suggests that affective polarization surrounding gun control is predominantly driven by event occurrence, reflecting immediate emotional and reactive discourse patterns rather than pre-event ideological anticipation.
%
%Similar to our previous analysis, we categorized all conversations into three distinct polarization levels based on their extremity of affective polarization: low ($x < 2.5$), moderate ($2.5 \leq x \leq 7.5$), and high ($x > 7.5$). Figure~\ref{fig:res_gc_percentage} displays the proportion of conversations falling into each category before, during, and after the event. From Figure~\ref{fig:res_gc_percentage}, we observe that extreme polarization ($x > 7.5$) is substantially higher during and after mass shooting events across all cases. Unlike our past observations in climate change events—where pre-event polarization spikes were observed in anticipation of scheduled events—gun control discussions primarily exhibit a reactive polarization pattern. This means polarization levels spike in direct response to mass shootings rather than as a result of pre-existing ideological debates. Our analysis highlights that affective polarization in gun control discourse is event-driven and shows a distinct reactive pattern in nature in contrast to climate change discussions, where polarization trends were structured around anticipated events.

%%%%%%%%%%%%%%%%%%%%%%%%%%%%%%%%%%%%%%%%%%%%%%%%%%%%%%%%%%%%%%%%%%%
\section{Discussion}
\label{sec:con}
In this study, we introduced a novel framework combining large language models (LLMs) and predefined heuristics to quantify affective polarization in online discussions on charged topics like climate change and gun control. Our hybrid approach uses LLMs to efficiently extract stance, affect, and agreement dynamics at scale, while domain experts guide the polarization scoring process through intuitive rules. The primary goal is not benchmarking LLM performance, but presenting a structured, human-in-the-loop framework that accelerates affective polarization analysis.

% In this study, we introduced a novel framework leveraging large language models (LLMs) and expert-defined heuristics to quantify affective polarization in online discussions for socially and politically charged topics like climate change and gun control. Our method provides an intuitive way to integrate LLMs with human-in-the-loop expertise, where LLMs excel in extracting stance, affect, and agreement dynamics at scale, while domain expert-defined rules quantify the degree of affective polarization in large-scale social media conversations. This hybrid approach enables structured and automated computation of affective polarization scores, allowing LLMs to handle the labor-intensive stance, affect, and agreement identification, while the critical task of scoring is guided by domain experts. We want to note that the primary goal of this study is not to benchmark LLM performance but to propose a novel human-in-the-loop framework that combines LLM-based annotation with domain expert heuristics to accelerate and structure affective polarization analysis.

Our results show that climate change and gun control events significantly influence affective polarization. Climate discussions exhibited anticipation-driven polarization, with extreme opinions emerging before events like the IPCC report release or the UN Environment Assembly. Gun control debates showed a reactive pattern, with polarization intensifying primarily after mass shootings. The volume of conversations also surged following these events, highlighting real-world triggers in online discourse. Although most conversations remained moderately polarized, spikes in extreme polarization aligned closely with major events, providing deeper insights into the temporal dynamics of affective polarization.

% Our results demonstrate that both climate change and gun control events significantly influence the intensity and distribution of affective polarization. Climate change discussions exhibited anticipation-driven polarization, where extreme opinions often surfaced before predefined events like the IPCC report release or the UN Environment Assembly. In contrast, gun control debates displayed a reactive pattern, with extreme polarization intensifying primarily during and after mass shooting incidents. The number of conversations also surged in the aftermath of these events, further reinforcing the role of real-world triggers in shaping online discourse. Moreover, through comparative analysis, we observed that while the majority of conversations generally fall within the moderate polarization range, spikes in extreme polarization align closely with major events. These findings provide deeper insights into the temporal nature of affective polarization in social media conversations. 

Future work could extend this analysis by including a broader range of events and topics, leveraging the generalizability of our proposed framework. Validating LLM-based polarization scoring across diverse sociopolitical contexts and exploring interventions to mitigate extreme polarization could also offer valuable insights for policymaking and social media governance. In summary, our study highlights the benefits of combining AI-driven analysis with domain informed frameworks, enabling interpretable measurement of online polarization, even in brief interactions.

% Future work could extend this analysis by incorporating a broader range of events, as the proposed framework is generalizable and can be adapted to additional topics. Further validation of LLM-based polarization scoring across different sociopolitical contexts, along with research into strategies for mitigating extreme polarization, such as counter-narrative interventions could provide valuable implications for social media governance and policymaking. To summarize, our study underscores the importance of combining advanced AI-driven stance and sentiment analysis with expert-defined frameworks to develop interpretable measures of online polarization even in small conversations consisting of single interactions. By bridging computational and domain expertise, this approach paves the way for more informed discussions on the impact of social media narratives on public opinion and policy debates.

%%%%%%%%%%%%%%%%%%%%%%%%%%%%%%%%%%%%%%%%%%%%%%%%%%%%%%%%%%%%%%%%%%%
\section*{Acknowledgment}
This research was supported by the Army Research Office under Grant W911NF-22-1-0035. The views expressed are those of the authors and do not necessarily reflect the official policies of the Army Research Office or the U.S. Government. The U.S. Government retains the right to reproduce and distribute reprints for governmental purposes.

\bibliographystyle{splncs03}
\bibliography{reference}

\end{document}